\documentclass{article}
\usepackage{spconf,amsmath,graphicx}

\usepackage{mwe} % to get dummy images
\usepackage{comment}
\usepackage{multirow}
\usepackage{colortbl}
\definecolor{mygray}{gray}{.9}
\newcolumntype{L}{>{$}l<{$}} 
\usepackage[T1]{fontenc}

\title{Analyzing the Spatiotemporal Interaction and Propagation of ATN Biomarkers in Alzheimer's Disease Using Longitudinal Neuroimaging data}

\name{Qing Liu $^{\dagger}$ \quad Defu Yang $^{\star}$ \quad Jingwen Zhang $^{\ddagger}$ \quad Ziming Wei $^{\ddagger}$  \quad Guorong Wu$^{\star}$ \quad Minghan Chen$^{\ddagger}$\thanks{Correspondence: chenm@wfu.edu}}

\address{
$^{\dagger}$Wake Forest University, Department of Mathematics and Statistics, Winston-Salem, NC 27109 \\
$^{\star}$University of North Carolina at Chapel Hill, Department of Psychiatry, Chapel Hill, NC 27599\\
$^{\ddagger}$Wake Forest University, Department of Computer Science, Winston-Salem, NC 27109}

\begin{document}
% \graphicspath{{./fig/}}

\maketitle
\begin{abstract}
Three major biomarkers: $\beta$-amyloid (A), pathologic tau (T), and neurodegeneration (N), are recognized as valid proxies for neuropathologic changes of Alzheimer's disease.
While there are extensive studies on cerebrospinal fluids biomarkers (amyloid, tau), the spatial propagation pattern across brain is missing and their interactive mechanisms with neurodegeneration are still unclear.
To this end, we aim to analyze the spatiotemporal associations between ATN biomarkers using large-scale neuroimaging data.
We first investigate the temporal appearances of amyloid plaques, tau tangles, and neuronal loss by modeling the longitudinal transition trajectories. Second, we propose linear mixed-effects models to quantify the pathological interactions and propagation of ATN biomarkers at each brain region. 
Our analysis of the current data shows that there exists a temporal latency in the build-up of amyloid to the onset of tau pathology and neurodegeneration.
The propagation pattern of amyloid can be characterized by its diffusion along the topological brain network.
Our models provide sufficient evidence that the progression of pathological tau and neurodegeneration share a strong regional association, which is different from amyloid.
\end{abstract}

% -------------------------------------------------------------------------
\section{Introduction}
\label{sec:intro}
% Jingwen - please follow this structure
% One short paragraph (2-3 sentences): General info of AD 
% One long paragraph: describe both temporal of ATN and Spatial of ATN
% One paragraph: our hypothesis and planned work

Alzheimer's disease (AD) is a chronic neurodegenerative disorder pathologically characterized by $\beta$-amyloid deposition (A), pathologic tau (T), and neurodegeneration (N)~\cite{jack2018nia}.
The spread of tau and neurodegeneration are more alike both in spatial and temporal scales, while amyloid exhibits a distinct progression pattern~\cite{arriagada1992neurofibrillary}. 
The accumulation of amyloid, commonly seen as the earliest sign of AD, could start decades before significant neurodegeneration is detected. Tau, considered as the downstream effect of amyloid, is closely related to neurodegeneration chronologically. 
The spatial progression pattern also varies between A and T/N biomarkers. While senile plaque appears first and aggregates mostly in precuneus and frontal lobe, neurodegeneration is commonly detected and developed near entorhinal cortex and hippocampal area~\cite{musiek2015three}. 
Studies show that aggregated, phosphorylated tau occurs first near the brain stem entorhinal cortex and hippocampus within the temporal lobe; tau will spread out to neocortex area, in most case, only with the presence of abnormal amyloid deposition. %~\cite{elobeid2012hyperphosphorylated}.

%by amyloid plaques, neurofibrillary tangles, and neurodegenerations. 
However, due to the complexity of the underlying mechanisms, the detailed disease progression remains uncertain, rendering a gap between current neuropsychological assessments and clinical diagnoses.
With the improvement of neuroimaging technology, the AD field has witnessed the inclusion of non-invasive imaging biomarkers in the diagnosis scheme, which provide essential spatial information for AD study. 
The longitudinal amyloid, tau, and FDG positron emission tomography (PET) scans, along with diffusion-weighted images (DWI), allow us to investigate the progression of AD in a systematic and comprehensive way. 
In this paper, we will (1) study the longitudinal trajectories of ATN biomarker transitions using logistic regression; (2) propose linear mixed-effects models to characterize the topological spread of amyloid across the brain and (3) examine the pathological interactions of ATN biomarkers. 
Our results offer adequate evidence in testing hypotheses about the spatiotemporal evolution of ATN biomarkers in Alzheimer's disease.

\section{Materials and Methods}
\label{sec:method}
\subsection{Subjects and data}

From the ADNI database, subjects underwent longitudinal scans of PET and DWI are eligible for the present study. The analysis of this paper is based on a significant amount of processed neuroimaging data that includes  2753 amyloid PET scans from 1204 subjects, 839 tau PET scans from 539 subjects, 3478 FDG PET from 1449 subjects, and 504 DWI from 141 subjects. Specifically, each DWI scan was parcellated into 148 cortical regions using the Destrieux atlas~\cite{destrieux2010automatic} and formulated to a 148$\times$148 connectivity matrix by applying the tractography in FreeSurfer. %~\cite{reuter2012within}. 
Standardized uptake value ratio (SUVR) was calculated for amyloid, tau, and FDG PET at each brain region (node). 

%Mild Cognitive Impairment (MCI) is a disorder that signifies the transitional stage between age-related cognitive decline and Alzheimer disease~\cite{petersen2004mild}, and is applied in the disease stage classification. 
Each subject's scan has a diagnostic label, categorizing its dementia stage as one of: cognitive normal (CN), significant memory concern (SMC), early mild cognitive impairment (EMCI), late mild cognitive impairment (LMCI), and AD. 
%Since there is hardly a clear observed difference in biomarker levels between CN, SMC, and EMCI groups, 
We group subjects into two cohorts based on their baseline scans: the normal cognition cohort (NC) that includes subjects labeled as CN, SMC, and EMCI; and the AD cohort (AD) that includes subjects labeled as LMCI and AD. 
 
\subsection{Statistical analysis}
\label{sec:statAnalysis}
All statistical analyses and modeling in this work are implemented using R statistical programming language. Table~\ref{tab:demographic} summarizes the baseline demographic characteristics of NC and AD cohorts, as well as the changes of amyloid, tau, and FDG between individuals' baseline and last scans. 
%Note that we measure the neurodegeneration by $\Delta N_{ij}= FDG_{i}-FDG_{j}$.
Note that FDG scans (a measure of cortical thickness) are used as the opposite indicator of neurodegeneration in this paper.
We also include the average network connectivity and spectral gap (of the Laplacian matrix) that characterizes the diffusion speed on brain networks. 

\vspace{-0.02\linewidth}
\begin{table}[h]
\centering
\caption{Demographic and biomarker characteristics.}
\label{tab:demographic}
\footnotesize{
\begin{tabular}{|lcc|}
\hline
\rowcolor{mygray}
\textbf{All subjects}  & \textbf{NC (752)} & \textbf{AD (636)}    \\
\hline
Age (year) & 72.57 $\pm$ 6.67  & 73.38 $\pm$	7.72  \\
Education (year) & 16.30 $\pm$	2.64 & 16.02$\pm$2.74   \\
Gender (female) & 40\%  &  48\%  \\
APOE4 (+) & 58\%   & 34\%  \\
% Marital (married) & 80\%  &  72\%  \\
%Ethnicity (NonHis) & 95\%  & 97\%   & \chi^2_2=1.47, .480\\
%Race (White)& 92\%   & 94\%   & \chi^2_6=6.37, .383\\
\hline
\rowcolor{mygray}
\textbf{Amyloid subjects} & \textbf{NC (547)} & \textbf{AD (214)}  \\ 
\hline
$\Delta A$ (node avg.) & 0.065 $\pm$	0.190 & 0.050 $\pm$	0.193   \\
$\Delta A$ (node 124)  &0.047$\pm$	0.217  & 0.031 $\pm$ 0.216  \\
\hline
\rowcolor{mygray}
\textbf{Tau subjects} & \textbf{APOE4 -- (86)} & \textbf{APOE4 + (60)}  \\
\hline
$\Delta T$ (avg.)   & 0.001 $\pm$ 0.075 & 0.048 $\pm$	0.070\\
$\Delta T$ (node 147)   & 0.026 $\pm$ 0.120 & 0.093 $\pm$	0.143\\
\hline
\rowcolor{mygray}
\textbf{FDG subjects} & \textbf{NC (749)} & \textbf{AD (631)} \\ 
\hline
$\Delta N$ (avg.) & 0.018 $\pm$ 0.054  &  0.035 $\pm$ 0.053  \\
$\Delta N$ (node 59) & 0.007 $\pm$ 0.088 & 0.043 $\pm$ 0.102  \\
\hline
\rowcolor{mygray}
\textbf{DWI subjects}  & \textbf{NC (94)} & \textbf{AD (47)}  \\ 
\hline
Connectivity (avg.) &  2.563 $\pm$ 0.081 & 2.532 $\pm$ 0.072    \\
Spectral gap (avg.) &  0.155 $\pm$ 0.056 & 0.134 $\pm$ 0.046   \\
\hline
\end{tabular}
}
\end{table}
\vspace{-0.02\linewidth}

In this paper, we propose three \textbf{hypotheses} to be validated on the large-scale longitudinal neuroimaging data. 
\textit{(1) Temporal discrepancy of ATN biomarkers.} We hypothesize that there is a time latency between the abnormal accumulation of amyloid, the onset of tau, and the onset of neurodegeneration.
\textit{(2) Topological spread of amyloid.} We hypothesize that amyloid protein diffuses along the brain structural network and diffusion rate follows the white matter connectivity strength between brain regions. 
\textit{(3) Pathological interaction of ATN biomarkers.} We seek to provide evidence that could support the interactive pathway of mainstream amyloid cascade hypothesis where amyloid activates the hyperphosphorylation process of tau protein, which then causes neurodegeneration and eventually leads to AD.

\textit{Characterizing the temporal discrepancy of ATN biomarkers.}
%Amyloid [A] was identified to precede and accelerate tau pathology [T] that could potentially lead to neurodegeneration [N]. To this end, we studied the longitudinal changes in A/T/N biomarker profiles.
To investigate the transition time from the build-up of amyloid to the onset of tau pathology and neurodegeneration, we first categorize the longitudinal biomarker burden for each subject by using receiver operating characteristic~\cite{landau2010comparing}.
% By performing the receiver operating characteristic (ROC) analysis on NC and AD cohorts, we search for the optimal cutoffs to classify subjects' longitudinal scans as normal (--) or abnormal (+)~\cite{landau2010comparing}. 
We search for the optimal cutoffs to classify subjects' longitudinal scans as normal (--) or abnormal (+).  The threshold that maximizes the sum of sensitivity and specificity assuming normally distributed data in both cohorts is chosen as the optimal cutoff value for each biomarker that is averaged across the brain.
%For simplicity, we used the averaged cortical thickness of ATN across the whole brain region from each subject in determining the cutoff values.
The cutoff values and the test accuracy of ATN biomarkers are: 1.97 and 77.35\% (A), 1.58 and 79.82\% (T), 0.82 and 74.16\% (N), respectively.
%And the corresponding sensitivity and specificity are:  64.73\% and 82.38\% (A),  44.29\% and 93.21\% (T), 74.26\% and 73.86\% (N). 
Provided with the categorized profiles, we then apply logistic regression to model the longitudinal trajectories of ATN biomarker transitions.

\textit{Characterizing the topological spread of amyloid.} 
Suppose $W$ is an adjacency matrix encoding the connectivity strength between brain regions and $M$ is a ``degree" matrix with each diagonal element denoting the total connectivity strength of the node. We can obtain the graph Laplacian matrix $L=M-W$ for each longitudinal scan of subjects. To quantify the influence of amyloid diffusion on its propagation pattern, we propose a linear mixed-effects model targeting the spatial and longitudinal change of amyloid burden,
\begin{equation}\label{model:lapA}
    \Delta A^{kn}_{ij} = \beta^{kn}_0+\beta_1 D^{kn}_{i}+\beta_2 \Delta t^{kn}_{ij}+\beta_3 DX^k + \epsilon^k_{ij},
\end{equation}
where $\Delta {A}^{kn}_{ij}={A}^{kn}_{j} - {A}^{kn}_{i}$ denotes the amyloid change between the $i$-th and $j$-th amyloid PET scans ($i<j$) of subject $k$ at node $n$ ($n=1,2,\dots,148$). 
Specifically, we include the predictor $D^{kn}_{i}$ as a proxy of amyloid diffusion between time $t_i$ and $t_j$, which is the $n$-th element of $D^{k}_{i}=-L^{k}_i * A^k_i$. Here $L^k_i$ is the Laplacian matrix of subject $k$ at time $t_i$ and $A^{k}_i$ is a vector denoting amyloid degrees across all brain nodes at time $t_i$. 
Considering the different amyloid levels between NC and AD dementia stages (denoted as $DX$) and the time span between two scans (denoted as $\Delta t_{ij}=t_j-t_i$), we include them as covariates in model \ref{model:lapA}.
The parameters $\{\beta_1, \beta_2,\beta_3\}$ are estimated by the restricted maximum likelihood approach and $p$-values are calculated using likelihood ratio test, here $\beta^{kn}_0$ is a subject-specific random intercept following a normal distribution $N(\mu_n, \sigma^2_{kn})$.
% while or here

\textit{Characterizing the pathological interaction of} \textit{ATN biomarkers.}
We propose the following linear mixed-effects models to investigate the possible pathological interactions of $(A, T)$ and $(T, N)$:
%Although tau PET is a relatively new modality, we would like to consider the possible mechanistic pathway that involve Tau and $A\beta$, and we considered the following linear regression to offer some insight on this matter. 
\begin{equation}\label{model:AT}
    \Delta T^{kn}_{ij} = \beta^{kn}_0+\beta_1 A^{kn}_{i}+\beta_2 \Delta t^{k}_{ij}+\beta_3 APOE^{k}   +\epsilon^k_{ij},
\end{equation}
\begin{equation}\label{model:TN}
    \Delta N^{kn}_{ij} = \beta^{kn}_0+\beta_1 T^{kn}_i + \beta_2\Delta t^{k}_{ij}+ \beta_3 DX^{k} +\epsilon^{k}_{ij}.
\end{equation}
% \begin{equation}\label{model:AN}
%     \Delta N^{kn}_{ij} = \beta^{kn}_0+\beta_1 A^{kn}_i + \beta_2\Delta t^{k}_{ij}+ \beta_3 APOE^{k} +\epsilon^{k}_{ij}
% \end{equation}
Here, $\Delta T^{kn}_{ij}$ and $\Delta N^{kn}_{ij}$ are the longitudinal changes of tau and neurodegeneration between the $i$-th to $j$-th scans of subject $k$ at node $n$, respectively. 
In model~\ref{model:AT}, we use the amyloid burden at time $t_i$ to explain the change of tau between the $i$-th and $j$-th scans.
We consider model~\ref{model:TN} to investigate whether tau is significantly associated with the longitudinal change of neurodegeneration. 
The estimated coefficients with $p$-values are reported in Table ~\ref{tab:result}.
% -----------------------------------------------------------------------
\section{Results and Discussions}

\subsection{Temporal latency in $A\rightarrow N$ and $A\rightarrow T$ transitions}
\begin{figure}[ht]
\centering
\includegraphics[width=0.95\linewidth]{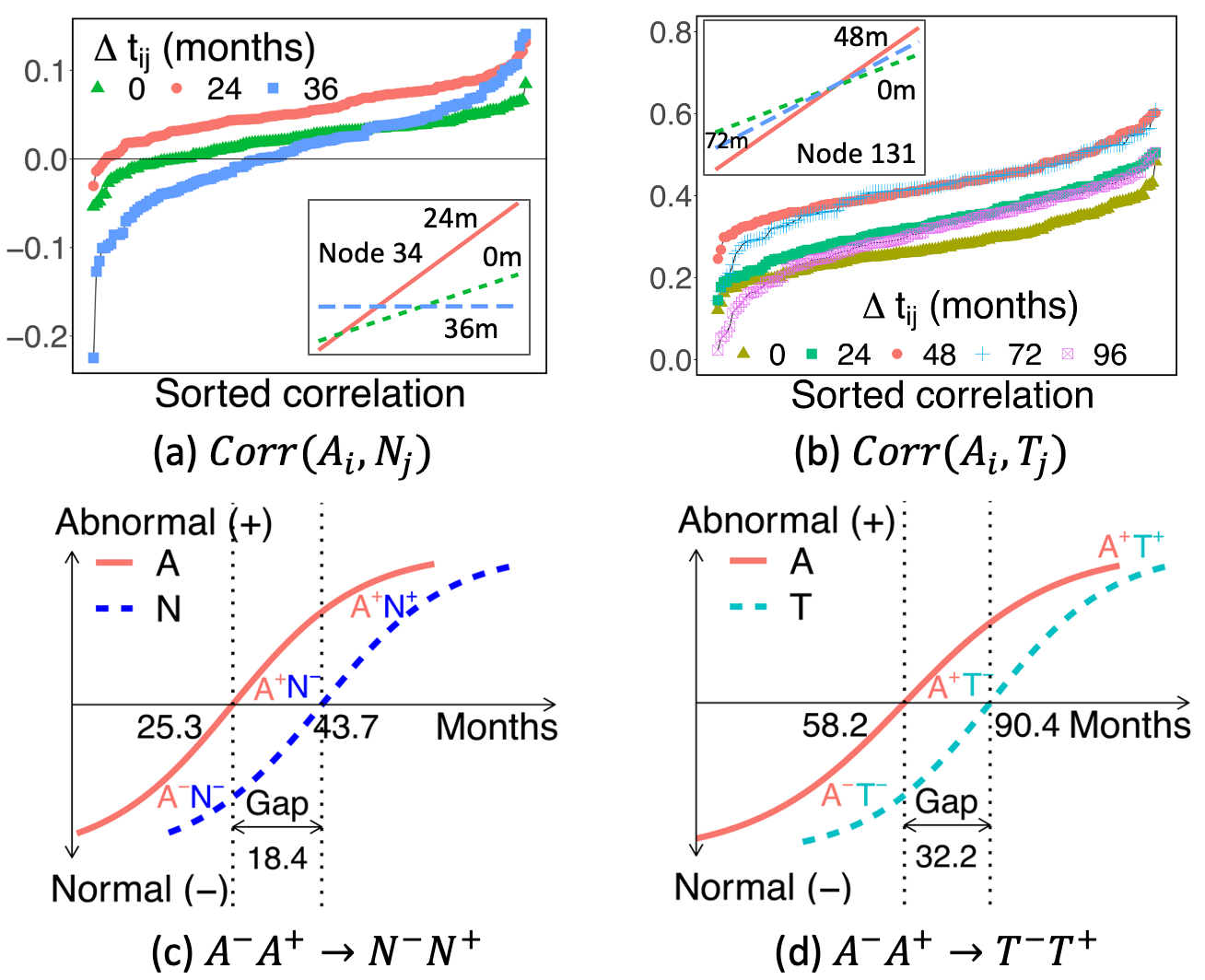}
\caption{\small{\textbf{(a, b)}
Pearson correlations of ATN biomarkers considering time span ($\Delta t_{ij}$) from 0, 24, 36, 48, 72, 96 months. Curves represent the correlations of $(A_i, N_j)$ and $(A_i, T_j)$ across all brain nodes, sorted in ascending order. The linear regression lines on nodes 34 and 131 are given in the embedded window as examples. \textbf{(c, d)} Longitudinal trajectories of biomarker transitions from normal (-) to abnormal (+) using cutoff values. 
%Subjects with the temporal evolution of $A \rightarrow N$ or $A \rightarrow T$ are selected for logistic fitting.
}}
\label{fig:sequence}
\end{figure}

As the temporal appearance of ATN biomarkers is a frequent contention, we study their correlations by dividing PET scans into different groups of pairs based on the time span between the two biomarkers. The correlations at all 148 nodes are ordered and presented in Fig.~\ref{fig:sequence}ab.
% Since the high amyloid burden could potentially trigger neurodegeneration, we see a large number of nodes showing a positive association. 
We see that a large number of nodes showing positive correlations as high amyloid burden could potentially trigger neurodegeneration.
When the time differences between amyloid and FDG scans are around 24$\pm 3$ months, we observe the strongest positive association between $A_i$ and $N_j$ (the red dotted curve in Fig.~\ref{fig:sequence}a). Taking node 34 as an example, the correlation has the steepest slope when $\Delta t_{ij}=24$ months.
Similarly, we see that the positive correlation between amyloid and tau reaches the highest level when the time differences between $A_i$ and $T_j$ are around 48$\pm 3$ months (the red dotted curve in Fig.~\ref{fig:sequence}b). The correlation remains almost the same for $\Delta t_{ij}=72$, then stabilizes around the level of $\Delta t_{ij}=24$.

We further analyze the longitudinal trajectories of $A\rightarrow N$ and $A\rightarrow T$ sequences using logistic regression with the optimal cutoff value found in Section~\ref{sec:statAnalysis}.
In Fig.~\ref{fig:sequence}c, we found that the average time lapse from $A^-A^+$ to $N^-N^+$ transitions could be as long as 18 months, where the red and blue curves denote the predicted trajectory of amyloid and neurodegeneration level. Meanwhile, Fig.~\ref{fig:sequence}d shows an average time lapse of 32 months from $A^-A^+$ to $T^-T^+$ transitions. It is worth noting that the 18-month gap between AN biomarkers roughly matches the 24-month that shows the strongest correlation, and the 32-month gap between AT biomarkers is close to the 48-month that has the highest association.   
The results in Fig.~\ref{fig:sequence} provide strong evidence to the hypothesis that amyloid plaques, tau tangles, and neuronal loss happen in a temporal discordant manner.

\subsection{Amyloid propagation on a whole-brain scale}

Applying model~\ref{model:lapA} to each node, we find that the model is effective for 47 nodes where all covariates are significant ($p<0.05$). 
In Table~\ref{tab:result}, the network diffusion shows a significant positive correlation to the longitudinal change of amyloid burden. The linear fitting of $D$ vs. $\Delta A$ on node 124 is given as an example shown at the bottom of Fig.~\ref{fig:brain}a. 
%The coefficient of $\Delta t$ is a small positive number, we may conclude that a longer period of time between two amyloid PET scans will result in a greater amount of average change in amyloid aggregation. 
The negative coefficient of DX indicates that the AD cohort on average has a relatively smaller change in amyloid compared to the NC cohort, which is consistent with our summary of $\Delta A$ in Table~\ref{tab:demographic}. 
One possible explanation is that subjects in the AD group with high accumulation of amyloid plaque, tend to stabilize at the current level and the further increment is limited to the brain volume. 
Other covariates, i.e., APOE4 and gender, are found to be insignificant and thus are not included in this model.

We see that the 47 nodes distribute across all five brain lobes, indicating that amyloid progresses along the fiber pathways (the grey edges in Fig.~\ref{fig:brain}a) on a whole-brain scale. 
This aligns with previous findings on the prion-like diffusion characteristic of amyloid~\cite{musiek2015three}. 
It is worth noting that in the evaluation of both tau and neurodegeneration propagation, their associations with network diffusion are insignificant at most nodes, which provides additional support for the hypothesis that amyloid protein diffuses and propagates via the brain structural network.

\subsection{Regional interactions of ATN biomarkers}
%Model 1 is fitted using different subsets of data, and we find that a maximum of six months time difference between $A_i$ and $T_i$ yields stabled results in the sense that the 20 nodes remain unchanged when selecting the nodes with all covariates being significant.

For models 2-3, we apply a filtering mechanism on our dataset (downsizing via time gap) to exclude the influence of possible outliers on model results. 
With this, we could obtain a stable set of nodes at which the models are effective and are not sensitive to data resampling randomness. 
% With this, we could obtain a stable set of nodes that have significant covariates and are not sensitive to data resampling randomness. 
 
\begin{figure}[ht]
\centering
\includegraphics[width=0.95\linewidth]{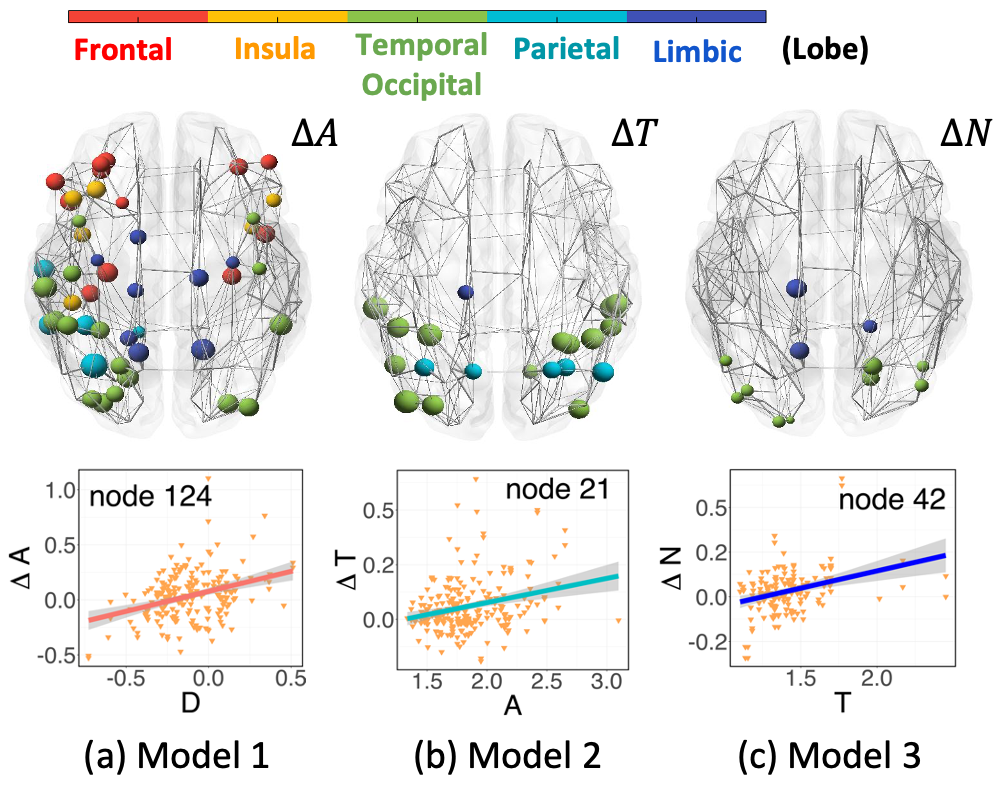}
\caption{\small{Row one: the plot of brain nodes with significant covariates from each model, where the node color denotes the located brain lobe and the size denotes the average longitudinal change of the corresponding biomarker at that node. The grey edge represents the node-to-node connectivity.
%Brain images were generated via BrainNet Viewer [46]. 
Row two: linear regression fittings of $D$ vs. $\Delta A$, $A$ vs. $\Delta T$, $T$ vs. $\Delta N$ in models 1-3.}}
\label{fig:brain}
\end{figure}

%% model 2
In model~\ref{model:AT}, 
amyloid burden is positively associated with the longitudinal change of tau, as shown in Table~\ref{tab:result} and Fig.~\ref{fig:brain}b (the linear regression plot for node 21).
Meanwhile, we observe a positive association between APOE4 and $\Delta T$, suggesting that APOE4 carriers have a higher risk of tau aggregation than non-carriers, which is also reflected in Table~\ref{tab:demographic} ($\Delta T$). 
In Fig.~\ref{fig:brain}b, our model~\ref{model:AT} is effective at 20 nodes and most of them fall into temporal and occipital lobes (green nodes) and parietal lobe (sky blue nodes). 
This clear pattern suggests that tau deposition primarily happens in the above three lobes.
Different from the result in model~\ref{model:lapA} where amyloid propagates across the cerebral cortex, model~\ref{model:AT} provides evidence that tau exhibits a distinct pattern and tends to specifically accumulate in clinically affected areas~\cite{ossenkoppele2016tau}.
%Contrary to model~\ref{model:lapA}, positive $\beta_3$ for dementia stage suggests that AD cohorts have a relatively larger change in neurodegeneration compared to NC cohort. 

%% model 3 & 4
Our result from model~\ref{model:TN} shows that the observed longitudinal neurodegeneration change is positively associated with the aggregation of tau (see Table \ref{tab:result} and the bottom plot of Fig.~\ref{fig:brain}c).
% We find 12 nodes located in the temporal, occipital, and limbic lobes, where the vision and memory with sensations are processed. 
%20 are from the temporal and occipital lobes, which suggests that tau PET could be effectively evaluated and treated as a biomarker of neurodegeneration, which is the proximate cause of cognitive decline.
% For model~\ref{model:AN}, 36 nodes scattered across the brain (Fig.~\ref{fig:brain}d) show significant positive correlations between amyloid burden and neurodegeneration, and could be used as predictive key nodes for assessing neurodegeneration in the early diagnosis of AD.
%% Summarize 2-3
% Overall, we observe positive associations between ATN biomarkers.
Specifically, models 2 and 3 share a similar spatial pattern with nodes mostly located in temporal, occipital, and limbic lobes, supporting the mainstream standpoint that there exists a strong regional association between the progression of pathological tau and neurodegeneration. 
%On the other hand, we also find nodes from frontal and insula lobes that present a positive association between the amyloid burden and neurodegeneration change.
%, which include 7 of the nodes we find from model~\ref{model:TN}, and others are largely coinciding with nodes from model~\ref{model:AT}, suggesting that tau's aggregation requires large amount of abnormal amyloid deposition.
Overall, we observe positive associations between ATN biomarkers (we applied the same type of model and confirmed the positive association between $A$ and $\Delta N$, which is not shown here).
Although both amyloid and tau are considered biomarker profiles of AD and most dementia is multifactorial, a model that considers the mixed effect of amyloid and tau to the neurodegeneration change does not yield expected outcomes based on our study of the current data.
%Thus is not reported in this paper.

\vspace{-0.02\linewidth}
\section{Conclusion}
\vspace{-0.02\linewidth}
In this paper, we studied the spatiotemporal interaction and propagation of ATN biomarkers using large-scale neuroimaging data from the ADNI database. In testing the proposed hypotheses, we found that
(1) there are significant time gaps between evolutions of amyloid plaques, tau tangles, and neuronal loss;
(2) amyloid propagates on a whole-brain scale and the longitudinal change of amyloid is strongly associated with the diffusion characterized by the Laplacian matrix of brain networks;
(3) progressions of pathological tau and neurodegeneration share a strong regional association in temporal and occipital areas that are largely affected in AD. 
In the future, with the inclusion of new neuroimaging data, we plan to reassess our findings and consider models that include the mutual effect of amyloid and tau on neurodegeneration.

\begin{table}[t]
\centering
\caption{Estimated coefficients and $p$-values. Only three nodes of each model are listed here.}
\label{tab:result}
\footnotesize{
\begin{tabular}{|cccc|}
\hline
\rowcolor{mygray}
\textbf{Model 1 (47)} &  $\Delta A \leftrightarrow D$ & $\Delta A \leftrightarrow$ DX  & $\Delta A \leftrightarrow \Delta t$ 	 \\
\hline
node 50&	.271 (<.001)	& -.115	(.002)	& .0020	(.008)\\
node 124&	.499 (<.001)	& -.096	(.014)	& .0026	(.002)\\
node 131&	.281 (<.001)	& -.086	(.008)	& .0024	(.002)\\
\hline
\rowcolor{mygray}
\textbf{Model 2 (20)} &  $\Delta T \leftrightarrow A$ & $\Delta T \leftrightarrow$ APOE &$\Delta T\leftrightarrow \Delta t$ 	 \\
\hline
node 21&	.081	(.005)&	.034 (.043)	& .0019	(.003) \\
node 58&	.078	(.006)&	.057 (.002)	& .0021	(.001)\\
node 131&	.068	(.004)&	.047 (.005)	& .0015	(.002)\\
\hline
\rowcolor{mygray}
\textbf{Model 3 (12)} & $\Delta N \leftrightarrow T$  & $\Delta N \leftrightarrow$ DX  & $\Delta N \leftrightarrow \Delta t$	 \\
\hline
node 42&	.166	(<.001) &	.039 (.033) &	.0011 (.013) \\
node 59&	.083	(.001) &	.038 (.048) &	.0010 (.042) \\
node 133&	.058	(<.001) &	.040 (.001) &	.0008 (.008) \\
\hline
% \rowcolor{mygray}
% \textbf{Model 4 (36)} & $\Delta N \leftrightarrow A$ & $\Delta N \leftrightarrow$ APOE & $\Delta N \leftrightarrow \Delta t$  \\
% \hline
% n56&	.040	(<.001)&	.011	(.033)&	.0013	(.033) \\
% n130&	.037	(<.001)&	.011	(.028)&	.0010	(.028) \\
% n146& 	.035	(<.001)&	.014	(.004)&	.0006	(.004) \\
% \hline
\end{tabular}
}
\end{table}
% \vspace{-0.04\linewidth}

\vspace{-0.03\linewidth}
\section{Compliance with Ethical Standards}
\vspace{-0.03\linewidth}
This is a numerical simulation study for which no ethical approval was required.
\vspace{-0.03\linewidth}
\section{Acknowledgments}
\vspace{-0.02\linewidth}
All authors have no conflict of interest to report.

% -------------------------------------------------------------------------
\vspace{-0.03\linewidth}
\bibliographystyle{IEEEbib}
\bibliography{refs}

\begin{thebibliography}{1}

\bibitem{jack2018nia}
C.R. Jack~Jr, D.A. Bennett, K.~Blennow, M.C. Carrillo, B.~Dunn, et~al.,
\newblock ``Nia-aa research framework: toward a biological definition of
  alzheimer's disease,''
\newblock {\em Alzheimer's \& Dementia}, vol. 14, no. 4, pp. 535--562, 2018.

\bibitem{arriagada1992neurofibrillary}
P.V. Arriagada, J.H. Growdon, E.T. Hedley-Whyte, and B.T. Hyman,
\newblock ``Neurofibrillary tangles but not senile plaques parallel duration
  and severity of alzheimer's disease,''
\newblock {\em Neurology}, vol. 42, no. 3, pp. 631--631, 1992.

\bibitem{musiek2015three}
E.S. Musiek and D.M. Holtzman,
\newblock ``Three dimensions of the amyloid hypothesis: time, space
  and'wingmen',''
\newblock {\em Nature neuroscience}, vol. 18, no. 6, pp. 800--806, 2015.

\bibitem{destrieux2010automatic}
C.~Destrieux, B.~Fischl, A.~Dale, and E.~Halgren,
\newblock ``Automatic parcellation of human cortical gyri and sulci using
  standard anatomical nomenclature,''
\newblock {\em Neuroimage}, vol. 53, no. 1, pp. 1--15, 2010.

\bibitem{landau2010comparing}
S.M. Landau, D.~Harvey, C.M. Madison, E.M. Reiman, N.L. Foster, P.S. Aisen,
  et~al.,
\newblock ``Comparing predictors of conversion and decline in mild cognitive
  impairment,''
\newblock {\em Neurology}, vol. 75, no. 3, pp. 230--238, 2010.

\bibitem{ossenkoppele2016tau}
R.~Ossenkoppele, D.R. Schonhaut, M.~Sch{\"o}ll, S.N. Lockhart, N.~Ayakta,
  et~al.,
\newblock ``Tau pet patterns mirror clinical and neuroanatomical variability in
  alzheimer’s disease,''
\newblock {\em Brain}, vol. 139, no. 5, pp. 1551--1567, 2016.

\end{thebibliography}
% }

\end{document}